\documentclass[manuscript]{acmart}

\usepackage{graphicx}
\usepackage{subcaption} 
\usepackage{xcolor}
\definecolor{deepsoftgreen}{RGB}{52, 116, 79}
\AtBeginDocument{%
  }

\setcopyright{acmlicensed}
\copyrightyear{2018}
\acmYear{2018}
\acmDOI{XXXXXXX.XXXXXXX}
\acmConference[Conference acronym 'XX]{Make sure to enter the correct
  conference title from your rights confirmation email}{June 03--05,
  2018}{Woodstock, NY}
\acmISBN{978-1-4503-XXXX-X/2018/06}

\begin{document}
\newcommand{\revised}[1]{\textcolor{blue}{#1}}
\newcommand{\yiliu}[1]{{\small\textcolor{purple}{\bf [YL: #1]}}}
\newcommand{\hangyue}[1]{{\small\textcolor{deepsoftgreen}{\bf [HY: #1]}}}
\newcommand{\yun}[1]{{\small\textcolor{blue}{\bf [yun: #1]}}}
\title{RiskLens: Human-Informed Incident Transfer for AI Risk Evaluation}



\author{Yiliu Tang}
\email{yiliut2@illinois.edu}
\affiliation{
    \institution{University of Illinois Urbana-Champaign}
    \department{School of Information Sciences}
    \city{Champaign}
    \state{Illinois}
    \country{USA}
}

\author{Hangyue Zhang}
\email{hz85@illinois.edu}
\affiliation{
    \institution{University of Illinois Urbana-Champaign}
    \department{School of Information Sciences}
    \city{Champaign}
    \state{Illinois}
    \country{USA}
}

\author{Yun Huang}
\email{yunhuang@illinois.edu}
\affiliation{
    \institution{University of Illinois Urbana-Champaign}
    \department{School of Information Sciences}
    \city{Champaign}
    \state{Illinois}
    \country{USA}
}

\renewcommand{\shortauthors}{Tang et al.}





\begin{abstract}

As AI capabilities advance, early-stage teams need better ways to judge whether historical incidents transfer to new products, demanding support beyond static risk checklists or one-off ideation tools. We present RiskLens, a human–AI workflow with an interactive system built around a live crawl, refreshable ranker, and extensible, human-informed schema. RiskLens analyzes 2,034 AI incidents and 10,443 AI products to rank transferable incidents using a feature schema spanning technical, operational, and user-facing attributes and supports an interface that presents ranked incident evidence, generated scenarios, risk metrics, and mitigation suggestions. In a study with 21 AI-familiar participants, users surfaced evaluation criteria that we added as 13 human-informed dimensions, including harm categories, failure mechanisms, and vulnerable groups. The feature-based model outperformed a generic baseline with an 86.2\% win rate, and remained stable on a twenty-month future holdout; after human-informed schema expansion, the updated model outperformed the same baseline on 91.4\% of pairs.

\end{abstract}


\begin{teaserfigure}
    \includegraphics[width=\textwidth]{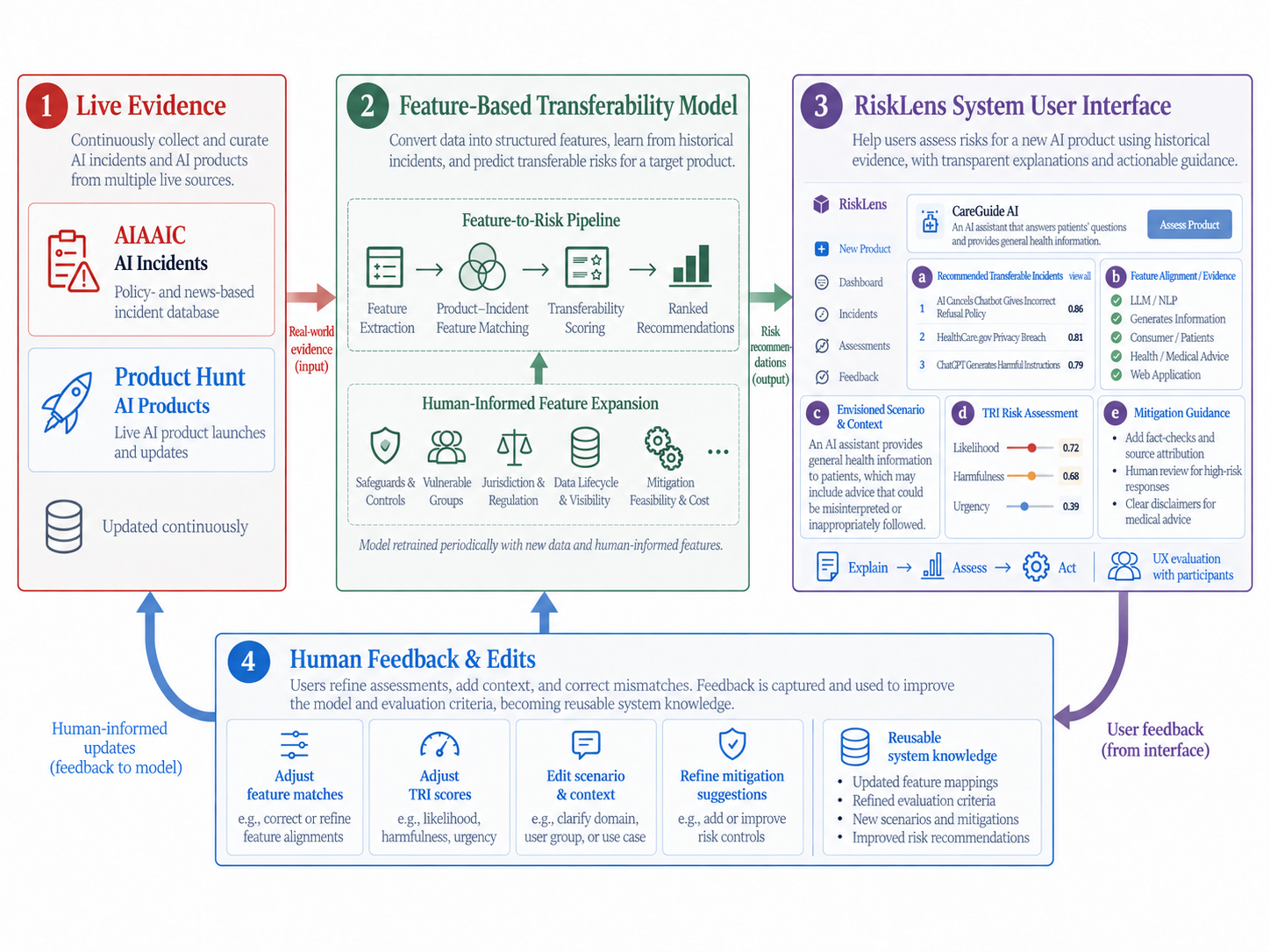}
    \caption{RiskLens co-evolving framework for AI risk evaluation. (1) Continuously updated evidence from AIAAIC incidents and Product Hunt products is transformed into structured features. (2) A feature-based transferability model matches products with historical incidents, scores their relevance, and produces ranked recommendations; human-informed features such as safeguards, vulnerable groups, jurisdiction, data lifecycle, and mitigation feasibility further expand the feature space. (3) The RiskLens interface presents ranked transferable incidents, feature alignment and evidence, envisioned scenarios, TRI risk metrics, and mitigation guidance; we also evaluate the interface through a user study. (4) Users refine feature matches, TRI scores, scenarios, and mitigations, and these edits are reintegrated as reusable knowledge to improve future recommendations and evaluation criteria.}
    \label{fig:incident}
    \Description{A four-part framework diagram. Panel 1, Live Evidence, shows two continuously updated sources: AIAAIC AI incidents and Product Hunt AI products. An arrow labeled real-world evidence leads to Panel 2, the Feature-Based Transferability Model, which shows a pipeline of feature extraction, product-incident feature matching, transferability scoring, and ranked recommendations, fed by a human-informed feature expansion row with safeguards and controls, vulnerable groups, jurisdiction and regulation, data lifecycle and visibility, and mitigation feasibility and cost. An arrow labeled risk recommendations leads to Panel 3, the RiskLens user interface, which shows a product, ranked transferable incidents with relevance scores, feature alignment evidence, an envisioned scenario, TRI risk sliders for likelihood, harmfulness, and urgency, and mitigation guidance. Panel 4, Human Feedback and Edits, sits below and shows users adjusting feature matches, adjusting TRI scores, editing scenarios and context, and refining mitigation suggestions, which become reusable system knowledge. Curved arrows return user feedback from the interface to Panel 4 and human-informed updates from Panel 4 back to the data and model panels.}
\end{teaserfigure}


\maketitle


\section{Introduction}

Anticipating potential harms is an important part of designing and
deploying AI products. Responsible-AI frameworks provide guidance
for identifying, assessing, and managing risks throughout
development~\cite{microsoft2022raistandard,nist_ai_risk_management}.
Historical incidents offer concrete evidence of how failures have
occurred in practice. Applying that evidence to a new product,
however, requires judging which conditions are shared and which
differences limit its relevance. For early-stage teams, the
challenge is to connect previous failures to plausible risks in
a product whose design and deployment context may still be
taking shape.

Existing resources support this assessment in several ways. The
AI Incident Database (AIID) and AIAAIC document incidents and
controversies that enable learning from previous
failures~\cite{mcgregor2020preventingrepeatedrealworld,
abercrombie2024collaborativehumancentredtaxonomyai}. The MIT AI Risk
Repository synthesizes risks from published frameworks and
classifications into a structured resource for identifying and
organizing concerns~\cite{slattery2025airiskrepositorycomprehensive}.
Design-stage tools such as Farsight, AHA!, ExploreGen, and BLIP
further help users explore potential uses, stakeholders, and
undesirable consequences through incident evidence or generated
examples~\cite{Farsight,buccinca2023aha,herdel2024exploregen,pang2024blip}.

Building on these capabilities, we investigate \emph{incident
transferability}: whether and why a failure documented in one
context offers useful evidence for assessing risks in another.
Products may share a technology or application domain while
differing in affected users, data practices, or safeguards.
Conversely, an incident from a different domain may reveal a
failure mechanism relevant to the target product. These
possibilities raise two connected questions: how can
product--incident comparison support the recommendation of
informative historical cases, and how can users' judgments reveal
additional criteria for assessing their applicability?

We present \textbf{RiskLens}, a human--AI collaborative workflow
and interactive system for early-stage risk assessment of new or
hypothetical AI products. A learned ranking model combines explicit
technical, operational, and user-facing feature matches with
textual similarity to recommend historical incidents. The
interface presents these incidents alongside feature alignments,
generated product-specific scenarios, Triple Risk (TRI) metrics
(likelihood, harmfulness, and urgency), and mitigation suggestions.
Users can examine the evidence and provide feedback on its
relevance and implications. This feedback provides material for
investigating both \emph{what to consider} when comparing incidents
with products and \emph{how to judge} whether the comparisons are
useful for risk assessment.

The evidence and criteria used in these comparisons may change
over time. Newly documented incidents expand the available cases,
while users' contextual knowledge can reveal considerations
missing from the initial representation. RiskLens therefore
supports routine corpus updates and an extensible feature schema.
Monthly ingestion adds products and incidents without retraining
the model; separately, our study-informed update expands the
schema and retrains the ranker. Across two dataset versions, the
corpus comprises 2,034 incidents and 10,443 products.

Our evaluation proceeds in three phases.
\textbf{Phase~1} evaluates the original feature-based model against
a generic baseline using an LLM-based pairwise comparison
protocol, complemented by a sampled human-agreement check. The
model was preferred in 86.2\% of comparisons and retained an
advantage on later products from a twenty-month temporal holdout,
with an 85.3\% win rate.
\textbf{Phase~2} examines how 21 AI-familiar participants use
RiskLens to assess a selected product. Participants' diagrams and
verbal accounts indicated broader consideration of risks, greater
use of incident evidence to justify judgments, and more structured
assessment processes. Their feedback also identified contextual
considerations missing from the initial representation.
\textbf{Phase~3} incorporates these findings into 13 additional
feature dimensions, regenerates product and incident
representations, and retrains the model. We evaluate the expanded
representation through ablation and compare its recommendations
with the generic baseline under human-informed evaluation
criteria. In this comparison, the expanded model was preferred
in 91.4\% of pairs. We also examine how the revised criteria appear
in the judge's rationales.

This paper makes three contributions:
\begin{itemize}
    \item \textbf{An interactive workflow for incident-based
    risk assessment.}
    RiskLens connects structured incident ranking with an
    interface for examining historical evidence, exploring
    product-specific scenarios, assessing risks, and considering
    mitigations.

    \item \textbf{An evaluated feature-based approach to
    incident recommendation.}
    We examine the utility of explicit product--incident
    attributes through comparison with a generic baseline,
    a temporal holdout evaluation, and ablation of a human-informed
    extension to the representation.

    \item \textbf{Empirical insights into users' assessment
    processes and transferability criteria.}
    A 21-participant study characterizes changes in users'
    articulated risk-assessment approaches and identifies
    considerations for \emph{what to compare} and \emph{how to
    judge relevance}. These findings inform 13 additional feature
    dimensions and revised evaluation criteria, connecting users'
    contextual reasoning to subsequent incident recommendation
    and assessment.
\end{itemize}

\section{Related Work}

\subsection{AI Risk Frameworks and Incident Evidence}

Responsible AI research provides resources for documenting,
assessing, and managing AI risks. Model Cards support documentation of
model characteristics, intended uses, and evaluation
results~\cite{Mitchell2019modelcard}; Fairlearn and SageMaker Clarify
provide tools for assessing fairness, bias, and model
behavior~\cite{bird2020fairlearn,Hardt2021SageMaker}; and the NIST AI
Risk Management Framework organizes risk-management activities across
the AI lifecycle~\cite{nist_ai_risk_management}. These resources establish
practices for examining AI systems and their potential impacts, while
risk taxonomies provide vocabularies for describing the concerns that
assessments should consider \cite{tang2026empowered}. For example, the MIT AI Risk Repository
synthesizes published risk frameworks into a database of risks organized
through causal and domain taxonomies~\cite{slattery2025airiskrepositorycomprehensive}.

Incident repositories offer a complementary form of evidence: accounts
of failures and controversies involving systems used in practice. The
AI Incident Database collects reports of realized harms and near harms
to support learning from previous failures~\cite{mcgregor2020preventingrepeatedrealworld},
while AIAAIC documents incidents and controversies involving AI,
algorithms, and automation~\cite{abercrombie2024collaborativehumancentredtaxonomyai}. Research on incident
documentation also considers how this evidence should be represented.
Agarwal and Nene propose a standardized incident schema that records
system characteristics, affected parties, causes, severity, and
harms~\cite{Agarwal_2024}. Such work establishes the importance of
structured descriptions for organizing and comparing incident evidence.
RiskLens addresses a related assessment task: using structured
descriptions of both incidents and target products to recommend
historical cases that may inform a new product's risk assessment.

\subsection{Tools for Anticipating Harms During AI Design}

HCI research has developed tools that help people anticipate consequences
before deploying a technology. AHA! uses language models to generate
stakeholder-specific examples of harms for AI deployment scenarios,
supporting reflection on how system failures may affect different
people~\cite{buccinca2023aha}. ExploreGen generates possible uses of AI
technologies and classifies their associated risk levels using the EU AI
Act's risk categories, supporting early-stage exploration by developers
and compliance experts~\cite{herdel2024exploregen}. These approaches help
make potential uses and consequences concrete enough for designers to
examine.

Other systems connect this anticipatory work to documented experiences.
BLIP extracts, summarizes, and categorizes undesirable consequences of
digital technologies from online articles, allowing researchers to
explore consequences beyond those they already know and identify
implications for their projects~\cite{pang2024blip}. Farsight integrates
harm exploration into AI application prototyping: it retrieves incident
reports using embedding similarity to a user's prompt and supports
interactive exploration and editing of generated use cases,
stakeholders, and harms~\cite{Farsight}. Together, these systems
establish the value of combining concrete accounts of harm with
opportunities for contextual reflection and exploration.

RiskLens builds on these systems by examining how explicit technical, operational, and user-facing attributes can inform the ranking of historical incidents for a target product.

\subsection{Human-Informed Criteria for Risk Assessment}

Research on AI design and evaluation shows why assessment criteria need to
account for the contexts in which systems are developed and used \cite{zheng2022ux, tang2025llmxr, tang2024metrics}.
Holstein et al. identify mismatches between the support offered by
fairness research and the needs of industry
practitioners~\cite{Holstein_2019}. Rakova et al. examine how
organizational culture, structures, and work practices enable or
constrain responsible AI initiatives~\cite{Rakova_2021}. These
findings motivate tools that accommodate users' contextual knowledge
alongside predefined technical descriptions. For incident-based
assessment, this raises questions about which characteristics make a
previous failure informative and which differences in deployment
conditions limit its relevance.

Human-AI systems also provide precedents for helping users develop and
revise the concepts through which they analyze information. LLooM enables
analysts to inspect and steer human-interpretable concepts generated
from unstructured text, including refining concepts and introducing
criteria informed by domain knowledge~\cite{Lloom}. In evaluation planning,
EvAlignUX supports researchers in exploring literature-grounded UX
metrics and their relationships to potential research
outcomes~\cite{zheng2025evalign}. These systems illustrate how AI
assistance can support the articulation and revision of analytical
criteria, motivating attention to users' assessment processes as well as
the outputs a system produces.

Building on work that incorporates contextual knowledge and supports human refinement of analytical criteria, RiskLens investigates how users’ judgments about incident relevance can inform reusable criteria for comparing historical incidents with new products.

\section{RiskLens Workflow Design}
\label{sec:method}

The literature reviewed in Sections~1 and~2 motivates four design
goals for RiskLens: recommend historical incidents relevant to a
target product, support users in examining their applicability,
incorporate user feedback into subsequent assessments, and allow
the evidence base and comparison criteria to be updated over time.
These goals guide the ranking model, interactive interface, and
feedback integration described below.



\begin{itemize}
  \item[\textbf{DG1}] Recommend historical incidents to support
    assessment of risk transferability to a target product.
    \item[\textbf{DG2}] Help users reason about transferability by
    surfacing relevant features, generated scenarios, risk metrics,
    and mitigations.
    \item[\textbf{DG3}] Capture user feedback to inform revisions to
    recommendations and evaluation criteria.
  \item[\textbf{DG4}] Support updates to the incident database,
    feature schema, and ranking model without system redesign.
\end{itemize}


\subsection{Backend Implementation}
\label{sec:backend}

\subsubsection{Dataset Construction}
\label{sec:datasets}

We constructed a product--incident corpus using AI products from
ProductHunt\footnote{\url{https://www.producthunt.com/}} and incident
records from AIAAIC~\cite{abercrombie2024collaborativehumancentredtaxonomyai}.
Product records include names, descriptions, technology categories,
purpose classifications, and image URLs. Incident records include
descriptions, information about the systems involved, risk
categorizations, and reported impacts. We manually reviewed and extracted
structured metadata to support consistency across the collected records.

The corpus comprises an initial collection, Dataset~v1, and a subsequent
set of additions, referred to as v2 additions. Dataset~v1 covers
May~2023--October~2024 and contains 9,688 products and 346 incidents.
The v2 additions cover November~2024--June~2026 and contain 755 new
products and 1,688 newly documented incidents. Together, the two
collections contain 10,443 products and 2,034 incidents.
Table~\ref{tab:datasets} summarizes their sizes and periods.

The model construction described below uses
Dataset~v1 and the original eight-dimensional feature schema. The
temporal evaluation examines the original model's performance on
subsequently collected products. The human-informed extension incorporates additional feature dimensions identified through the user study.

\begin{table}[h]
\small
\caption{Dataset v1 vs.\ v2 comparison.}
\label{tab:datasets}
\begin{tabular}{llll}
\toprule
 & \textbf{v1} & \textbf{v2} & \textbf{Total} \\
\midrule
Products       & 9,688 & 755 new     & 10,443 \\
Incidents      & 346                       & 1,688 new   & 2,034  \\
Period         & May~2023 – Oct~2024       & Nov~2024 – Jun~2026 & — \\
\bottomrule
\end{tabular}
\end{table}

\subsubsection{Feature Generation}
\label{sec:features}
We developed the feature schema through an initial LLM-assisted
exploration followed by researcher-led structuring. GPT-4o first
extracted free-form capability labels from product and incident descriptions. These labels were consolidated through LLM-assisted
grouping and merging into a taxonomy of 127 AI capabilities. Building on this
exploration, the researchers designed eight categorical dimensions
with predefined value sets to support structured comparison between
products and incidents. We encoded these dimensions and their permitted values in JSON
schemas. For each product and incident, GPT-4o received the item's
name and description and assigned
exactly one permitted value to each dimension. This process produced a common
representation for comparing products and incidents.


As shown in Table~\ref{tab:features_core}, the original schema contains
eight dimensions organized into three groups. \emph{Technical} features
describe technology type (16 permitted values) and model class (19);
\emph{operational} features describe application domain (32), deployment
setting (23), and scale of operation (16); and \emph{user-facing}
features describe target user group (30), interaction modality (24),
and output type (23).

\begin{table*}[htb!]
\centering
\caption{Model Feature Categories}
\label{tab:features_core}
\begin{tabular}{llp{4cm}l}
\hline
\textbf{Category} & \textbf{Feature} & \textbf{Definition} & \textbf{Example} \\
\hline
Technical & Technology Type & Core AI technology used & LLM, computer\_vision \\
& Model Class & Specific architecture & transformer, CNN \\
Operational & Application Domain & Industry/domain context & healthcare, finance \\
& Deployment Setting & Technical deployment & web\_app, mobile\_app \\
& Scale of Operation & Operational scale & individual, enterprise \\
User-Facing & Target User Group & Primary user population & professionals, general public \\
& Interaction Modality & User interaction method & text, voice, touchscreen \\
& Output Type & System output category & generation, classification \\
\hline
\end{tabular}
\end{table*}


Following the user study, we expanded the schema with 13 additional dimensions, such as vulnerable groups affected, jurisdiction and regulatory regimes, existing safeguards, mitigation feasibility, data lifecycle characteristics, and risk propagation patterns.

\subsubsection{Auto-Crawl Pipeline (DG4)}
\label{sec:crawl}
RiskLens uses a monthly ingestion pipeline to add products and incidents
to the corpus. After ingestion, GPT-4o extracts features for the new records using the
schema. Newly added products become
available for exploration, and newly added incidents become candidates
for recommendation.





\subsubsection{Model Training}
\label{sec:model}

\textbf{Training pairs and relevance labels.}
We considered all combinations of v1 products and incidents as candidate
pairs and sampled 50,000 pairs, stratifying across semantic-similarity
strata to balance the sampled distribution. GPT-4o rated each pair on six
dimensions: logical coherence, factual accuracy, contextual relevance,
application domain match, practical implementability, and exploitability.
Each dimension received a score from 1 to 5.

We averaged the six ratings and normalized the result to $[0,1]$:
\begin{equation}
    y(P,I) = \frac{1}{4}\left(\frac{1}{6}
    \sum_{j=1}^{6}s_j(P,I)-1\right),
    \label{eq:relevance_target}
\end{equation}
where $s_j(P,I)$ is GPT-4o's rating of product--incident pair $(P,I)$
on dimension $j$. Thus, a pair receiving a rating of 1 on every dimension
has target 0, while a pair receiving 5 on every dimension has target 1.
These targets represent LLM-assessed relevance for risk assessment.

\textbf{Hybrid ranking model.}
The model uses two signals to assess incident relevance:
categorical feature matches and text similarity. For each feature
dimension $i$, we set $m_i(P,I)=1$ when the product and incident
have the same categorical value, and $0$ otherwise. We encode product descriptions and incident reports using
Sentence-BERT, then apply a shared learned projection before
calculating cosine similarity. This allows the text representations
to be adjusted during training using the relevance labels.

The model combines these signals into a relevance score:
\begin{equation}
    \hat{y}(P,I)=\sigma\left(
    \sum_{i=1}^{d}\beta_i m_i(P,I)
    +\beta_{\mathrm{text}}\operatorname{sim}_{\mathrm{text}}(P,I)+b
    \right),
    \label{eq:relevance_prediction}
\end{equation}
where $d=8$ for the original model. The learned coefficients
$\beta_i$ weight the categorical matches, while
$\beta_{\mathrm{text}}$ weights text similarity $\operatorname{sim}_{\mathrm{text}}(P,I)$. The parameter
$b$ is a learned bias, and the sigmoid function $\sigma$
maps the combined value to a score between 0 and 1.

\textbf{Training procedure.}
We randomly split the 50,000 labeled pairs at the pair level
into 40,000 training pairs and 10,000 validation pairs.
During training, we jointly adjusted the text projection network
and scoring coefficients to minimize the mean squared error
between predicted relevance scores and GPT-4o-assigned labels.
We used Adam with a learning rate of 0.001 and trained for
100 epochs.

\subsubsection{Feature-Guided Retrieval and LLM Risk Contextualization}
\label{sec:retrieval}

When a user submits a product for assessment, RiskLens follows three
steps.

\textbf{Step~1 — Feature extraction.} The product description is passed
to GPT-4o, which extracts a structured 8-dimension feature vector.

\textbf{Step~2 — Incident ranking.} The model scores each incident in the available corpus using its
categorical feature matches and textual similarity to the product. The
top-$k$ incidents are returned as recommendations, with $k=5$ by default.

\textbf{Step~3 — Risk contextualization.} The product information and recommended incidents are provided to an LLM
to generate explanations of their applicability, scenarios describing
how related failures might arise, and mitigation suggestions. These
outputs support users in interpreting the incident evidence within their
product context.

\subsection{Frontend Design}
\label{sec:frontend}
\begin{figure}
    \centering
    \includegraphics[width=\textwidth]{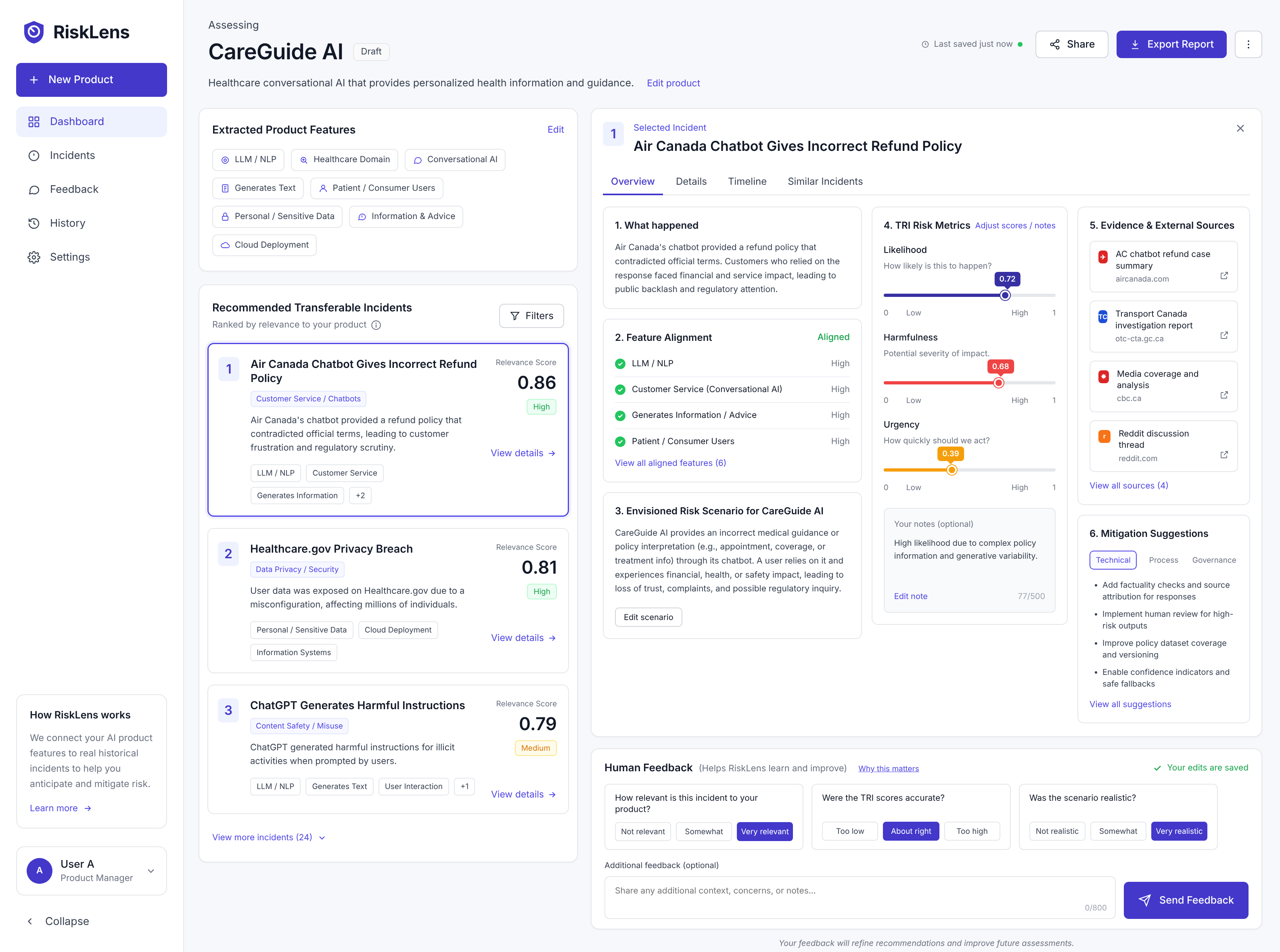}
    \caption{Example of RiskLens assessing a healthcare
    conversational AI product. A navigation sidebar appears on the left.
    The main workspace shows extracted product features and a ranked list
    of recommended incidents. An expanded incident view presents a
    summary, feature alignment, a generated risk scenario, likelihood,
    harmfulness and urgency ratings, external sources, and mitigation
    suggestions. Feedback collection below the incident view allows users
    to rate relevance, assess the risk ratings and scenario, and enter
    additional comments.}
    \label{fig:interface}
    \Description{Screenshot of the RiskLens interface assessing CareGuide AI, a healthcare conversational AI product. A left sidebar lists New Product, Dashboard, Incidents, Feedback, History, and Settings. The left column shows extracted product features as tags, such as LLM/NLP, Healthcare Domain, Conversational AI, and Personal/Sensitive Data, followed by three recommended transferable incidents with relevance scores of 0.86, 0.81, and 0.79. The right area shows the selected incident, Air Canada Chatbot Gives Incorrect Refund Policy, in six cards: what happened, feature alignment with four aligned features, an envisioned risk scenario for CareGuide AI, TRI risk metric sliders for likelihood (0.72), harmfulness (0.68), and urgency (0.39) with an optional note, evidence and external sources, and mitigation suggestions grouped into technical, process, and governance tabs. A Human Feedback panel at the bottom asks how relevant the incident is, whether the TRI scores were accurate, and whether the scenario was realistic, with a free-text field and a Send Feedback button.}
\end{figure}

The RiskLens incident review panel and the feedback collection interface are shown in Figure \ref{fig:interface}.

\subsubsection{Product Exploration and Ideation Panel (DG1)}

Users begin by defining a product or selecting an existing product from
the corpus. The product definition form accepts a name, description,
technologies, and purposes, with autocomplete suggestions drawn from
existing records. A searchable product grid allows users to browse
product descriptions and technology tags before selecting a product
for assessment.

\subsubsection{Incident Review Panel (DG1, DG2)}

The Incident Review Panel presents ranked recommendations and information
for assessing their applicability to the selected product.

\textbf{Incident evidence and feature alignment.}
Each incident card presents a description of what happened, its risk
domains, the model's relevance score, and extracted features. Alignment
badges highlight shared attributes between the product and incident.

\textbf{Generated scenarios.}
Product-specific scenarios describe how failures related to a historical
incident might arise in the selected product. These scenarios help users
examine the applicability of the incident to their own context.

\textbf{TRI risk metrics.}
Likelihood, harmfulness, and urgency provide initial assessments of the
potential risk in the target product. Likelihood expresses the assessed
plausibility of a similar failure, harmfulness concerns the severity of
its potential consequences, and urgency concerns how promptly the risk
should be addressed. Users can review and adjust these ratings based on
their understanding of the product and deployment context.

\textbf{Supporting sources.}
Links to original incidents, academic discussion and news coverage allow users to inspect the
underlying evidence and seek additional context.

\textbf{Mitigation suggestions.}
The interface presents LLM-generated suggestions informed by the
incident and product context, including technical controls and
organizational measures. Users can assess these suggestions for
relevance and feasibility when considering how to address a risk.

\subsubsection{Feedback Collection Interface (DG3)}

A chat-based interface elicits feedback on incident relevance, extracted
features, TRI ratings, generated scenarios, and mitigation suggestions.
The AI interviewer asks follow-up questions tailored to the selected
product and the user's responses. We analyzed this feedback to identify
additional comparison dimensions and assessment criteria.

\section{Evaluations}

We evaluated RiskLens through a system evaluation and a user study
with 21 participants. The system evaluation examined recommendation
performance and the contributions of individual features (RQ1).
The user study examined participants' risk assessment approaches
(RQ2) and elicited additional comparison dimensions and evaluation
criteria. We incorporated these findings into the system and
evaluated the resulting changes (RQ3). We investigated the following
research questions:


\textbf{RQ1}: \textit{How effectively does RiskLens recommend
historical incidents relevant to a target product's risk assessment,
and which features contribute most to these recommendations?}




\textbf{RQ2}:\textit{ How does RiskLens influence users' mental models for AI risk assessment?}


\textbf{RQ3}: \textit{How do users' judgments inform additional
comparison dimensions and evaluation criteria, and how do these
updates affect recommendation quality and evaluator rationales?}



The following subsections describe the user study. The system
evaluation methods are presented alongside their corresponding
results in Section~\ref{sec:results}.

\subsection{Participants}

The study was approved by the university's Institutional Review
Board. We recruited 21 participants through social media and
university networks. Participants included graduate students and
researchers with backgrounds in computer science, electrical and
computer engineering, information management, mathematics and
statistics, business, and the humanities. Twelve participants
(57.1\%) reported 3--5 years of experience in AI, machine learning,
or natural language processing; six (28.6\%) reported 1--2 years;
and three (14.3\%) reported less than one year. All participants
had hands-on AI experience during the preceding 12 months,
including prompting LLMs. Participants also reported experience
with classical machine learning and deep learning. Sessions lasted
1.5--2 hours, and participants received USD~20 per hour.


\subsection{Procedure}
The study began with participants completing an informed consent form and a pre-survey collecting background information. Researchers then introduced the study context and explained the tasks.
Participants were presented with the Product Exploration and Ideation Panel, where they spent approximately 10 minutes either selecting an existing AI product from the database or creating a new product by entering descriptions, technologies, and purposes. 

Before interacting with RiskLens' recommendations, participants used Figma to draw their initial mental model of AI risk assessment for their selected product. They engaged in think-aloud protocols to articulate their thought processes, capturing their baseline approach to identifying, categorizing, and prioritizing AI risks without system assistance. 

Participants then proceeded to the Incident Review Panel for their selected product, where researchers briefly introduced the system features. Participants explored incident cards, examining incident descriptions, extracted features, TRI metrics (likelihood, harmfulness, urgency), generated scenarios, external sources, and mitigation recommendations. For selected incidents of interest, participants interacted with the conversational feedback interface and provided feedback through the chat-based AI interviewer. 

After completing their interaction with RiskLens, participants returned to Figma to revise their mental model diagrams if they felt their risk assessment approach had changed. Finally, participants completed a post-survey evaluating their experience with RiskLens and participated in an exit interview to gather deeper insights about how the system influenced their risk assessment thinking, which features they found most valuable, and suggestions for improvement.

\subsection{Measures}

We collected mental model diagrams, survey responses, think-aloud
recordings, conversational feedback, and exit interviews to
examine participants' assessment approaches and experiences.


\subsubsection{Mental Model Diagrams.} 

Participants used Figma to represent their approach to assessing
risks for the selected product before and after interacting with RiskLens. We collected
these diagrams to examine changes in the risk considerations,
relationships, and assessment steps participants articulated.

\subsubsection{Survey.} The survey consisted of three main parts.
\textbf{Background information:} Participants answered questions about their AI literacy \cite{ng2022examination}, risk assessment familiarity, current practices, and professional background. 

\textbf{Perceptions of initial and revised risk assessment plans:} Participants rated their risk assessment plans before and after using RiskLens across 11 dimensions: novelty, relevance, efficiency, completeness, clarity, concreteness, quality, feasibility, rigor, risk considerations, and confidence. 

\textbf{System feature perceptions:} Participants rated each RiskLens feature (TRI metrics, scenarios, incidents, mitigations) across 8 Likert-scale items including usefulness, intention to use, cognitive load, inspiration, decision contribution, and planning contribution. Additional questions
examined perceived changes in participants' assessment approaches,
helpful combinations of features, and intentions to adopt or
recommend RiskLens.

\subsubsection{Think-Aloud Recordings.} We recorded participants' verbal reasoning while they drew initial mental models, explored incidents, and revised their mental models. This captured their decision-making processes and how they integrated RiskLens information into their risk assessment approaches.

\subsubsection{Interaction Records and Conversational Feedback}
The system recorded product inputs, incident exploration,
adjustments to TRI ratings, and feedback submissions. Through
the conversational interface, participants commented on incident
relevance, feature accuracy, risk ratings, scenarios, and
mitigation suggestions.

\subsubsection{Interviews.} We conducted exit interviews focusing on four areas: greatest value (most valuable aspects of using RiskLens), biggest limitation (most frustrating or limiting aspects), missing features (capabilities that would improve usefulness), and background impact (how professional experience shaped system perception and usage).

\subsection{Data Analysis}
We combined quantitative analysis of survey responses with
qualitative analysis of diagrams and verbal accounts to examine
changes in participants' assessment approaches and identify
criteria for refining RiskLens.

\textbf{Statistical Assessment of Survey Data.} 
We used Wilcoxon signed-rank tests to compare participants'
pre- and post-interaction plan ratings across the 11 dimensions.
We summarized feature perceptions and overall system perceptions
using descriptive statistics.


\textbf{Mental Model Analysis.} 
We compared participants' initial and post-interaction diagrams
to identify conceptual additions, process reorganizations, and
changes in how risks were categorized and connected.

\textbf{Think-Aloud and Interview Transcript Analysis.} 
Verbal protocols and interview data underwent iterative thematic coding by two researchers to extract insights about decision-making processes and system impact. This analysis specifically focused on identifying ``what to consider'' factors that participants found missing from current risk assessment approaches and ``how to judge'' criteria that influenced their evaluation of incident relevance and system recommendations.

\section{Results}
\label{sec:results}
We report findings in two parts. \textbf{RQ1} evaluates RiskLens as an \emph{internal system evaluation}: we compare recommendation models and examine what drives model performance. \textbf{RQ2} and \textbf{RQ3} report \emph{user study findings}: how participants used RiskLens in risk assessment work, and how human feedback shaped both what the system recommends and how it judges recommendations. We present RQ1 first to establish the evaluation details before analyzing user-facing outcomes.


\subsection{RQ1: RiskLens's Feature-Based Recommender Significantly Improves Recommendations}



\subsubsection{Hybrid Model Architecture and Feature Importance Analysis}

Table~\ref{tab:coefficients} reports the learned coefficients of the
hybrid model. Technology type received the largest categorical
coefficient ($\beta=1.178$), followed by application domain ($1.058$),
scale of operation ($0.940$), and deployment setting ($0.859$).
Because the categorical inputs are binary match indicators, these
coefficients describe the contribution of changing a mismatch to a
match while holding the other model inputs fixed. Technology type
therefore provides the strongest positive categorical match signal
within the trained model, followed by application domain, scale of operation, and deployment setting.


Output type ($\beta=0.762$), target user group ($0.650$), model class
($0.632$), and interaction modality ($0.489$) received smaller positive
weights. All categorical coefficients contribute to the score before
the sigmoid transformation; their values are not direct changes in
the final relevance score. Text similarity also received a positive
coefficient ($0.503$). We assess the contribution of text through
ablation comparisons, since its continuous values and learned
representation differ from the binary categorical inputs.


\vspace{0.5cm}
\begin{table}[h]
\centering
\caption{Learned Feature Coefficients in Full Model}
\label{tab:coefficients}
\begin{tabular}{lcc}
\hline
\textbf{Feature} & \textbf{Coefficient} & \textbf{Importance} \\
& $\beta$ (S.E.) & \textbf{Rank} \\
\hline
Technology Type & 1.178*** (0.011) & 1 \\
Application Domain & 1.058*** (0.010) & 2 \\
Scale of Operation & 0.940*** (0.010) & 3 \\
Deployment Setting & 0.859*** (0.010) & 4 \\
Output Type & 0.762*** (0.010) & 5 \\
Target User Group & 0.650*** (0.009) & 6 \\
Model Class & 0.632*** (0.010) & 7 \\
Interaction Modality & 0.489*** (0.010) & 8 \\
Text Similarity & 0.503*** (0.010) & — \\
\hline
\multicolumn{3}{l}{\footnotesize{*** $p < 0.001; ** p < 0.01; * p < 0.05$}} \\
\end{tabular}
\end{table}

We compared configurations using different combinations of categorical
features and text, measuring mean squared error (MSE) and mean absolute
error (MAE) against relevance labels on the validation
set. The four highest-weighted categorical features were technology
type, application domain, scale of operation, and deployment setting.
Table~\ref{tab:ablation} reports the results.

\begin{table*}[htb!]
\centering
\caption{Ablation Study. Impact of Feature Removal on Model Performance}
\label{tab:ablation}
\begin{tabular}{lcccc}
\hline
\textbf{Model Configuration} & \textbf{Validation Loss} & \textbf{MAE} & \textbf{Features Used} & \textbf{Performance} \\
& \textbf{(MSE)} & & & \textbf{Drop (\%)} \\
\hline
Full Model (All 8 + Text) & 0.0177 & 0.0732 & Tech, App, Scale, Deploy, & - \\
& & & Output, User, Model, Inter, Text & \\
Top 4 + Text & 0.0299 & 0.1193 & Tech, App, Scale, Deploy, Text & +68.9 \\
All 8 Only & 0.0336 & 0.1067 & Tech, App, Scale, Deploy, & +89.8 \\
& & & Output, User, Model, Inter & \\
Top 4 Only & 0.0336 & 0.1300 & Tech, App, Scale, Deploy & +89.8 \\
Tech + App + Text & 0.0349 & 0.1303 & Tech, App, Text & +97.2 \\
Tech + App Only & 0.0377 & 0.1356 & Tech, App & +113.0 \\
Tech Type + Text & 0.0415 & 0.1489 & Tech, Text & +134.5 \\
Text Only & 0.0569 & 0.1895 & Text & +221.5 \\
\hline
\multicolumn{5}{l}{\footnotesize{Performance drop calculated relative to Full Model validation loss}} \\
\end{tabular}
\end{table*}

The full model achieved the lowest MSE (0.0177) and MAE (0.0732).
Removing text increased MSE to 0.0336, an 89.8\% increase, indicating
that text contributed to relevance prediction even when all eight
categorical features were available. Retaining text with the top four
categorical features yielded an MSE of 0.0299. This configuration had
lower MSE than the eight-feature categorical-only configuration,
although its MAE was higher (0.1193 versus 0.1067).

The contribution of the remaining four categorical dimensions is
apparent when comparing the full model with Top 4 + Text: removing
these dimensions increased MSE by 68.9\%. Without text, the eight-feature
and four-feature configurations had the same MSE at the reported
precision, but retaining all eight features reduced MAE from 0.1300
to 0.1067. The text-only configuration had the highest MSE (0.0569)
and MAE (0.1895) among the evaluated configurations. Together, these
comparisons support combining categorical matches with learned text
similarity for predicting the relevance labels.

\subsubsection{Comparison with a Text-Similarity Baseline}
\textbf{Evaluation setup.}
We compared the feature-based model with a baseline that ranks incidents
by cosine similarity between Sentence-BERT embeddings of product and
incident descriptions. Both methods searched the full incident corpus
available for the evaluation. The 50 evaluation products did not appear
in the model's training data. For each product, we compared the two
methods' recommendations at corresponding positions in their top-five
lists. Excluding identical recommendations left 246 comparisons from
250 possible pairs.

GPT-4o evaluated which incident in each pair was more relevant to the
target product's risk profile using a structured rubric. The judge was
blinded to which method produced each recommendation. Training used
pointwise relevance labels, whereas this evaluation compared pairs of
recommendations for held-out products. GPT-4o was used in both stages.

\textbf{Recommendation preferences.}
The judge preferred the feature-based recommendation in 212 of 246
comparisons (86.2\%) and the text-similarity recommendation in the
remaining 34 (13.8\%). At the pooled comparison level, a proportion
z-test against a 50\% preference rate yielded $z=11.35$, $p<.001$.
Cohen's $h$, calculated relative to 0.5, was 0.81. 
The feature-based win rates were 85.7\%, 77.6\%, 92.0\%, 87.5\%, and
88.0\% at ranks 1 through 5, respectively. Thus, the judge preferred
the feature-based recommendations at each evaluated ranking position.


\textbf{LLM-as-Judge Human Agreement.}
LLM-based evaluators have increasingly been used for scalable pairwise and rubric-based evaluation~\cite{zheng2023mtbench,liu2023gevalnlgevaluationusing}, although prior work has identified validity concerns such as prompt sensitivity, position bias, self-preference, and divergence from human judgments~\cite{wang-etal-2024-large-language-models-fair, wei2025systematicevaluationllmasajudgellm, li-etal-2024-split, panickssery2024llmevaluatorsrecognizefavor, wataoka2025selfpreferencebiasllmasajudge, 10.1145/3708359.3712091}.
Accordingly, a necessary condition for interpreting our win-rate results is that the GPT-4o
judge's preferences align with human judgment. To validate this, we adopted a strategy similar to that in \cite{zheng2025evalign, pang2024blip} and randomly selected 50 pairwise comparisons from the evaluation set. One author determined which incident was more relevant to the product's risk profile using the same structured rubric presented to the LLM judge. A second author verified these assessments. Agreement between the human rater and GPT-4o was $\kappa = 0.71$
(86\% agreement, $n = 50$ pairs), indicating substantial agreement per
\citet{landis1977measurement} ($\kappa \geq 0.60$). These results suggest that GPT-4o's pairwise judgments were largely consistent with human assessments on the sampled comparisons.

\textbf{Why use a learned ranker rather than direct LLM recommendation?}
Using an LLM to compare two supplied incidents requires substantially
less input than providing the entire incident repository. With 2,034 incident records containing descriptions and structured
features, including the full corpus in a single recommendation
prompt would exceed the context window of the model available
when RiskLens was designed. 
LLM-based reranking of an initial shortlist remains a possible
extension.


\textbf{Examples favoring the feature-based recommendations.}
For Selzy, an email marketing tool, the judge preferred the
feature-based recommendation concerning researchers using ChatGPT
to generate convincing phishing emails that could bypass spam
filters. The baseline instead recommended an incident in which
Amazon Alexa suggested a dangerous physical challenge to a user.
The judge emphasized that \textit{``the incident directly involves
the generation of emails, which is central to Selzy's purpose''}
and that \textit{``the ability to generate phishing emails could
be exploited within Selzy's platform.''} This comparison illustrates
how shared application characteristics and output types can make
an incident relevant to a target product's potential misuse.

For Fryda, a digital wallet, the judge preferred the feature-based
recommendation concerning a Character AI incident in which users
could access other users' private conversations and account
information. The baseline recommended an incident concerning
Adobe Firefly's use of AI-generated images from competing systems
as training data, raising questions about training-data provenance.
The judge explained that a \textit{``privacy breach where users
access others' sensitive data is critical for a digital wallet
storing receipts and important information.''} This example
illustrates how an incident from a different application domain
can inform assessment of risks involving sensitive user information.

\textbf{An example favoring the text-similarity baseline.}
For Out Of The Blue, a B2B SaaS intelligence platform, the judge
preferred the baseline's recommendation concerning naviHealth's
use of AI to deny insurance coverage for post-acute care, including
cases in which patients reportedly still needed care. The
feature-based model instead recommended an incident concerning
ChatGPT's collection and use of conversation data for training
without explicit user consent. The judge emphasized the connection
between \textit{``automated systems making consequential decisions''}
and explained that \textit{``naviHealth's context of automated
decisions affecting critical outcomes is closer to Out Of The
Blue's operational needs for driving time-critical interventions.''}
This comparison illustrates a case in which the baseline surfaced
an operational connection that the judge considered more relevant
than the feature-based recommendation.

\subsubsection{Recommendation Performance on a Twenty-Month Temporal Holdout}

\textbf{Setup.}
To assess temporal robustness, we evaluated whether the feature-based model trained exclusively on Dataset~v1 (May~2023–Oct~2024; 9,688 products, 346~incidents) maintains its advantage when applied to Dataset~v2 (Nov~2024–Jun~2026; 755~new products, 1,688~newly documented incidents). We randomly sampled 150~v2 products for evaluation and ran the same LLM-as-comparator pairwise protocol used in the original v1 assessment.

\textbf{Feature taxonomy validity and change in feature frequencies.}
The existing extraction prompt assigned a non-null value to each of
the eight dimensions for all 755 v2 products, using the original
permitted categories. This demonstrates that the extraction procedure
could be applied to the later products without changing the schema;
assignment completeness alone does not establish the accuracy or
adequacy of those categories.

The frequencies of technology categories (\textit{tech\_type}) differed between the two
collections. LLMs accounted for 27.5\% of v1 products
($2,660/9,688$) and 31.9\% of v2 additions ($241/755$).
The share assigned to \texttt{autonomous\_agents} increased from
1.0\% ($100/9,688$) to 18.0\% ($136/755$), while
\texttt{rule\_based} declined from 24.7\% ($2,397/9,688$)
to 4.0\% ($30/755$).

\textbf{Temporal performance evaluation.}
Using the same pairwise evaluation protocol, the feature-based model was applied to 150~randomly sampled v2 products. For each product, we retrieved the top-1 incident using (a)~the feature-based model and (b)~the generic text-similarity baseline, then asked GPT-4o to judge which recommendation was more relevant to the product's risk profile. \textbf{The feature-based model won 128 out of 150 comparisons (85.3\%)}, compared to 86.2\% on v1 ($z=8.65$, $p<.001$, Cohen's $h=0.78$). The model retained an advantage over
text-similarity retrieval on the later products under this protocol.


\subsection{RQ2: Changes in Users' Risk Assessment Mental Models}



\begin{figure*}[ht!]
    \centering
    \includegraphics[width=\linewidth]{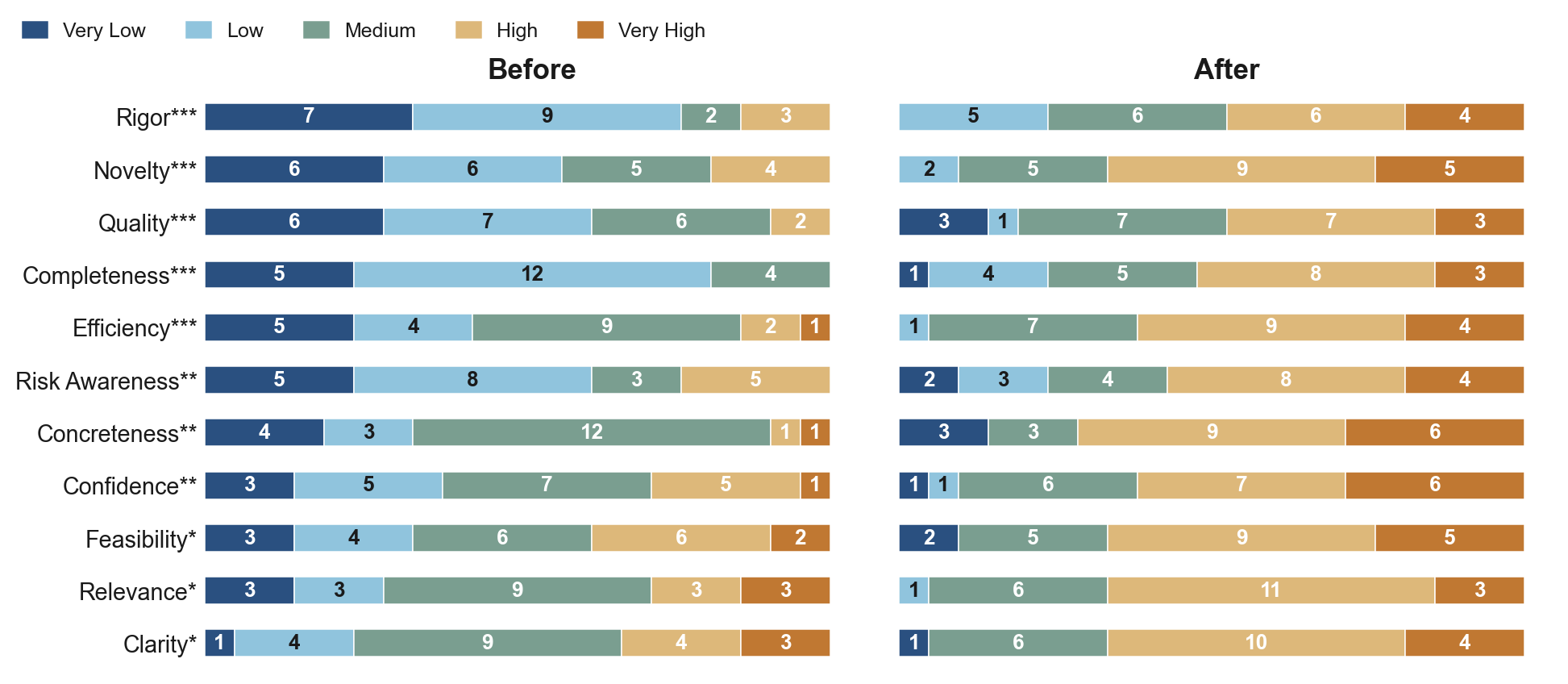}
    \caption{Distributions of participants' ratings of their initial
    (left) and revised (right) risk assessment plans. Revised plans
    received higher ratings across all 11 dimensions. (*: $p<.05$, **: $p<.01$, ***: $p<.001$; effect sizes $r_b$ = .47–.95.)}
    \label{fig:distribution}
    \Description{Two side-by-side stacked horizontal bar charts, labeled Before and After, showing how 21 participants rated their risk assessment plans on 11 dimensions using a five-level scale from very low to very high. Dimensions from top to bottom are Rigor, Novelty, Quality, Completeness, and Efficiency (each p < .001), Risk Awareness, Concreteness, and Confidence (each p < .01), and Feasibility, Relevance, and Clarity (each p < .05). In the Before chart, most ratings are very low, low, or medium. In the After chart, the bars shift toward high and very high on every dimension. For example, Rigor rises from 3 high or very high ratings before to 10 after.}
\end{figure*}

We examined participants' ratings of their initial and revised risk
assessment plans, their perceptions of system features, and changes
illustrated through mental model diagrams and verbal accounts. Across
the 21 participants, revised plans received higher ratings on all
11 dimensions, with unadjusted Wilcoxon signed-rank tests yielding
$p<.05$ for each comparison (Figure~\ref{fig:distribution}). The study
did not include a control condition; these findings describe
participants' assessments and reasoning following the study activities
and do not isolate the effect of the recommender.

\subsubsection{Self-Rated Plans and Perceived Decision Support}
\label{sec:risklens user result}


~\\

\textbf{Perceived Workflow Impact} 
Participants gave higher ratings of their revised plans on Efficiency
(W=0, p<.001, $r_b=1.00$) and Quality
(W=6, p<.001, $r_b=.91$). Participants also agreed that their revised mental models were better
overall (M=3.62, SD=1.28) and that their assessment approaches had
changed meaningfully (M=3.81, SD=1.12). In a separate question,
they anticipated that using RiskLens would make their assessment work
more time-efficient (M=3.86, SD=1.06). P4, for example, described
moving from not consciously assessing risks to developing an explicit
framework for doing so. These accounts suggest that participants
perceived benefits for both their plans and their assessment processes. Participants expressed interest in future use, reporting willingness
to recommend RiskLens to colleagues ($M=4.05$, $SD=0.80$) and to use it
in their work ($M=3.86$, $SD=0.91$).



\textbf{Perceived Decision Support} 
Revised plans received higher ratings for Clarity
(W=10, p=.017, $r_b=.74$) and Feasibility
(W=12, p=.017, $r_b=.74$). Participants also reported
greater Confidence in their plans (W=0, p=.002,
$r_b=1.00$) and higher ratings on the Risk Awareness item
(W=0, p=.002, $r_b=1.00$), which asked whether their
plans considered potential assessment limitations. These results
indicate that participants viewed their revised plans as clearer,
more feasible, and more attentive to their own limitations. In P2's
case, references to privacy and accuracy developed into a more
structured connection between user groups, use cases, and system
design considerations.

\subsubsection{Feature Use and Perceived Value}
\label{sec:combined feature}
~\\

Participants identified combinations of features as helpful to their
assessment process. Six participants (28.6\%) selected using all
features together as the most effective combination, while seven
(33.3\%) selected generated scenarios and mitigation suggestions.
Their ratings and explanations highlighted different contributions
from incident evidence, scenarios, TRI metrics, and mitigations.



\textbf{TRI metrics.}
Participants rated TRI metrics positively for Usefulness
(M=3.76, SD=1.09) and Decision Contribution
(M=3.76, SD=1.22). Their accounts emphasized support for deciding
which risks to address. As P3 and P6 noted, TRI metrics helped them
to \textit{``improve risk prioritization.''}


\textbf{Scenario generation.}
Generated scenarios received high ratings for Inspiration
(M=4.43, SD=0.93), Planning Contribution
(M=4.10, SD=0.77), and Ease of Understanding
(M=4.33, SD=0.91). Participants described scenarios as
\textit{``the most valuable aspect''} (P3),
\textit{``a great way to start putting all possibilities together''}
(P5), and \textit{``very helpful to envision otherwise abstract
risks''} (P6). Others described them as \textit{``useful''} (P11)
or \textit{``really clear helping to realize long-term effects''}
(P13). These accounts illustrate how concrete scenarios helped
participants explore potential consequences for their products.


\textbf{Real-world AI incidents.}
Historical incidents received high ratings for Intention to Use
(M=4.19, SD=0.98) and Assessment Contribution
(M=4.05, SD=1.07). Participants valued incident examples as inputs
for identifying risks and developing their assessment approaches.
P5 described them as \textit{``a great way for my brain to start
putting all the multitude of possibilities together.''}


\textbf{Mitigation measures.}
Participants rated mitigation suggestions positively for Ease of
Understanding (M=4.20, SD=0.70) and Planning Contribution
(M=4.25, SD=0.64). They described the suggestions as
\textit{``useful''} (P11) and \textit{``clear and planning-oriented''}
(P13), although some wanted more domain-specific detail (P5, P7).
These responses suggest that mitigations offered practical starting
points that participants could assess against their product contexts.


\subsubsection{Changes in Breadth, Depth, and Process Structure}
~\\
Participants' diagrams and verbal accounts illustrated changes in
the breadth of risks considered, the reasoning supporting their
judgments, and the organization of their assessment processes.
Separate post-survey items captured participants' perceptions of
these changes.

\begin{figure*}[ht!]
    \centering
    \includegraphics[width=\linewidth]{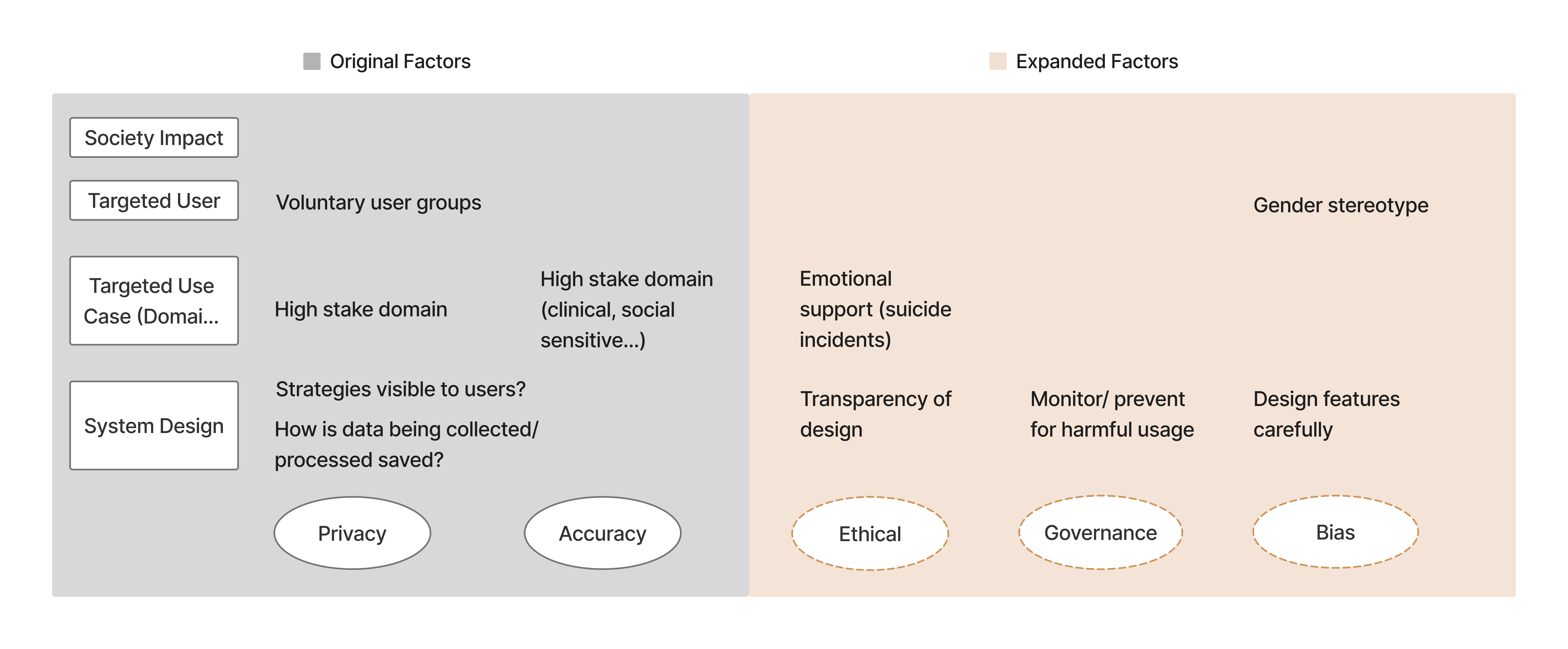}
    \caption{Changes in the breadth of P2's risk assessment
    considerations. The initial diagram appears on the left
    (grey); additions made after using RiskLens appear on the
    right (orange), including ethical, governance, and
    bias-related concerns.}
    \label{fig:p2prevspost}
    \Description{A diagram of P2's risk assessment considerations split into two regions. The grey left region, Original Factors, lists four row labels, Society Impact, Targeted User, Targeted Use Case (Domain), and System Design, with notes such as voluntary user groups, high-stake domains, whether strategies are visible to users, and how data is collected, processed, and saved, plus two risk ovals, Privacy and Accuracy. The orange right region, Expanded Factors, adds gender stereotypes, emotional support (suicide incidents), transparency of design, monitoring or preventing harmful usage, and designing features carefully, plus three dashed risk ovals, Ethical, Governance, and Bias.}
\end{figure*}

\textbf{Breadth of consideration.}
Participants reported considering a broader range of risk categories
(M=4.24, SD=0.83). Figure~\ref{fig:p2prevspost} illustrates P2's
expansion from an initial focus on technical risks to the inclusion
of ethical and governance concerns. This example shows how the
revised assessment incorporated considerations absent from the
initial diagram.


\begin{figure*}[ht!]
    \centering
    \includegraphics[width=\linewidth]{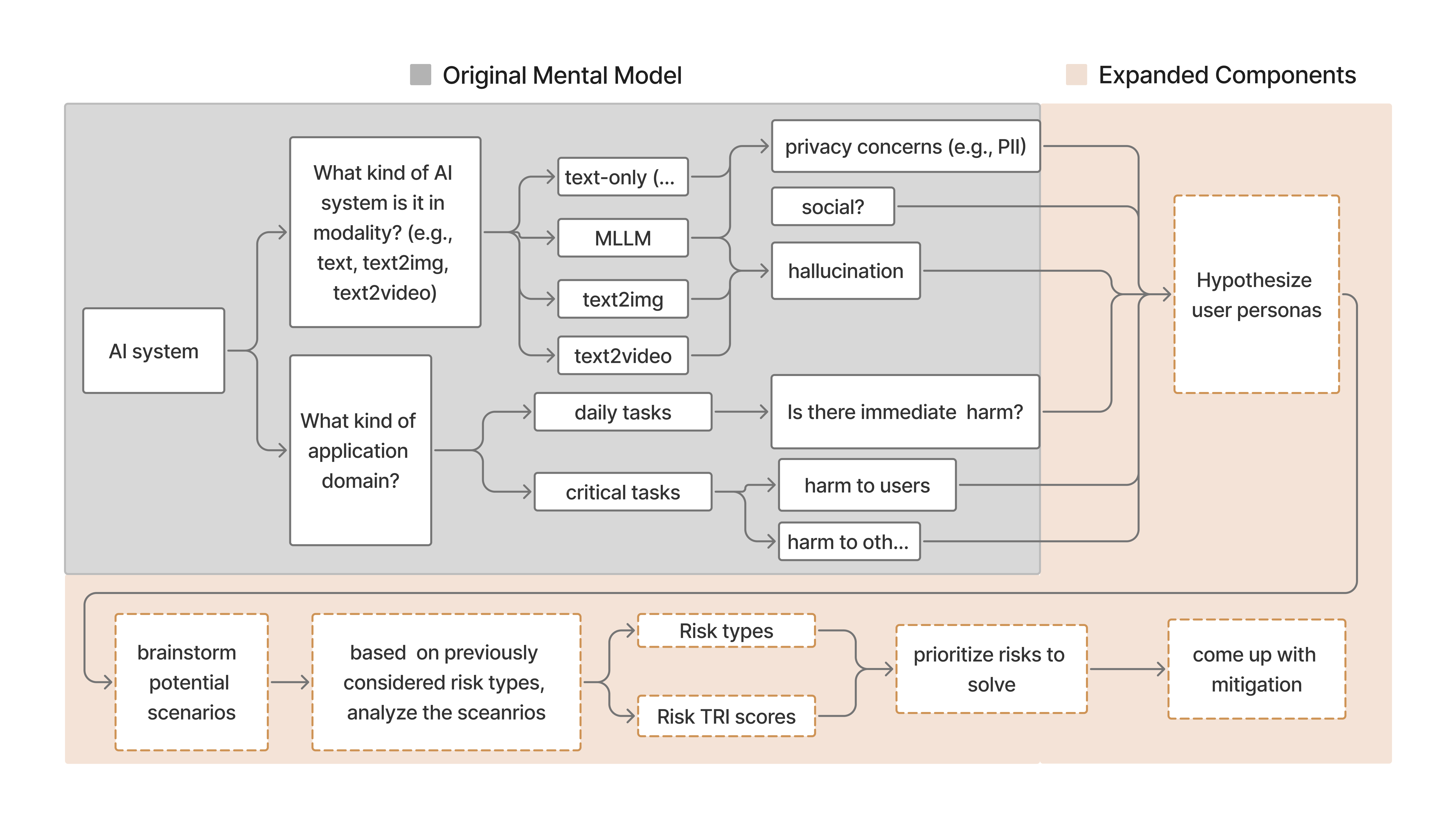}
    \caption{ Changes in the structure of P6's risk assessment
    process. The initial diagram appears at the top (grey);
    the revised process appears below (orange), incorporating
    scenario brainstorming, risk categorization, prioritization,
    and mitigation planning after using RiskLens.}
    \label{fig:p6prevspost}
    \Description{A flowchart of P6's risk assessment process. The grey region, Original Mental Model, branches from AI system into two questions: what kind of AI system it is in modality, which leads to text-only, MLLM, text2img, and text2video and then to privacy concerns, social concerns, and hallucination; and what kind of application domain it is, which leads to daily tasks and critical tasks and then to immediate harm, harm to users, and harm to others. The orange region, Expanded Components, connects all of these to a new step, hypothesize user personas, followed by a new sequence: brainstorm potential scenarios, analyze the scenarios based on previously considered risk types, assess risk types and TRI scores, prioritize risks to solve, and come up with mitigation.}
\end{figure*}

\textbf{Depth of reasoning.}
Participants reported greater reliance on evidence and data
(M=4.24, SD=0.89). Their explanations illustrated how incident
descriptions and sources helped them articulate reasons for their
judgments. P8 reflected, \textit{``at first my judgment was just
like a gut feeling, but then I looked at the descriptions and the
sources and I realized I could justify my rating with those. It
made my decision feel more defensible.''} P13 similarly explained,
\textit{``Initially I just wrote bias, misinformation, and job
displacement, kind of from my intuition. But after seeing these
incidents and scenarios, I started to think, okay, these are real
cases, so I should adjust my scores based on that evidence.''}
These accounts describe a shift toward using incident evidence
to explain assessment decisions.


\textbf{Structure of process.}
Participants reported thinking more systematically about AI risk
assessment (M=4.19, SD=0.93) and developing more sophisticated
assessment processes (M=4.24, SD=1.04). Ratings of improved
prioritization methods were more modest (M=3.52, SD=0.98).
Figure~\ref{fig:p6prevspost} illustrates P6's progression from
listing risks to a workflow that included scenario brainstorming,
TRI metrics, and explicit prioritization. Together with the verbal
accounts, this example illustrates how participants incorporated
elements of RiskLens into the structure of their assessment plans.



\subsection{RQ3: Human-Informed ``\textit{what to consider}'' and ``\textit{how to judge}''}



Through qualitative analysis of user-study transcripts, we identified
additional considerations participants brought to assessing historical
incidents and their applicability to new products. We organize these
findings around \emph{what to consider}, which informed extensions to
the product-incident representation, and \emph{how to judge}, which
informed revisions to the evaluation criteria. We then report the
integration and evaluation of these human-informed changes.


\subsubsection{What to consider}\label{sec:what}

\textbf{Vulnerable Groups.} Users emphasized the need to explicitly capture affected populations beyond end-users. As P2 and P7 noted during their pre-study mental modeling: \textit{``If there are like sensitive factors, for example, vulnerable user group...''}. Users wanted the system to distinguish between general public users and vulnerable populations (children, elderly, marginalized communities) through enhanced vulnerable\_groups\_affected, with additional consideration for community-level impacts.

\textbf{Jurisdiction \& Legal Context}. Multiple users highlighted the importance of regulatory landscape. P4 emphasized jurisdictional differences: \textit{``In Europe… we shouldn't use personal information… In the United States, it's not really clear… some company already received the informed consent… it wouldn't really apply to this context''.} P5 requested country-specific risk indicators: \textit{``copyright laws… change country to country''}. P4 noted: \textit{``Country matters. Cause legislation varies.''} and requested information about \textit{``What's the legal problem?''} These accounts motivated jurisdiction\_country fields and regulatory\_regimes tags to capture compliance requirements, with emphasis on how existing consent practices can make scenarios irrelevant in different regulatory contexts.

\textbf{Implementation Reality}. Users requested information about existing controls and feasibility constraints. P6 emphasized: \textit{``Mitigation measures: can point out the gap, from the current implementation towards the ideal case. Mitigation measures: can also point out the feasibility.''} P4 highlighted the dynamic nature of safeguards, noting that \textit{``user consent''} is a moving target as a product evolves and requested that the system consider \textit{``drawbacks and operational realities''}. This drove requirements for new schema fields including safeguards\_present, mitigation\_feasibility (low/mid/high), and mitigation\_cost\_hint to capture resource constraints and changing organizational contexts.

\textbf{Data Lifecycle}. A novel insight emerged around data handling practices. P8 noted during their exploration: \textit{``Your data, like, might be, uh, shared with companies without your consent... it's like not okay, like it's sensitive data, it's like, HIV diagnosis, stuff like that''}. This concern informed data\_types, data\_sources, and data\_visibility fields to capture collection, storage, and exposure patterns.

\textbf{Scenario Structure \& Propagation}. Users wanted to understand how risks spread and develop into concrete harms. P5 requested \textit{``a visualization/network of the scenario, from what goes wrong to how it spreads to the damage''} and suggested \textit{``Maybe have a timeline. A direction, a trend''}. This drove schema additions for spread\_mode (viral/contained) and spread\_velocity (fast/slow) to model risk diffusion patterns, combined with harm\_categories that specify concrete consequences.

\subsubsection{How to judge}\label{sec:how}
Participants' accounts also informed how we evaluated incident
relevance. We translated their concerns about contextual fit,
jurisdiction, affected populations, implementation, and expertise
into revised evaluation criteria.

\textbf{Contextual Fit}. Participants examined whether an incident's failure conditions
aligned with the target product's operation. P8 questioned the
connection between a privacy incident and its generated scenario,
observing, \textit{``The actual incident is that... [specific privacy violation]... And then, the scenario... What goes wrong? Is someone realizing they've been oversharing with AI?... So it's like, it feels a little bit disconnected''}. P3 wanted evaluation to reflect sensing differences (e.g., passive vs active capture), motivating checks of modality and data flow. This means judges should rate incidents in terms of operational characteristics (data flow, decision processes, user interaction patterns).

\textbf{Jurisdiction \& Legal Fit}. The judge should weigh country-specific rules and local consent practices when deciding relevance. P5 noted that some suggestions \textit{``might... only apply... for this country''}, while P4 provided a concrete counterfactual: if a company already has informed consent, \textit{``it wouldn't really apply to this context''}. This means the LLM judge should evaluate whether regulatory mismatches or existing compliance practices make an incident recommendation irrelevant to the target deployment context.

\textbf{Diverse Impact Assessment}. Participants wanted assessments to consider the people and groups
affected by a failure. P8 noted: \textit{``The damage can be more diverse, social damage, user damage, to what happens to society, what happens to me. Rather than just reputation damage''}. This concern informed criteria that examine individual,
organizational, and societal consequences relevant to the target
product. The emphasis is on identifying the applicable harms and
affected parties, rather than treating organizational consequences
as a sufficient account of impact.

\textbf{Implementation Focus}. Participants valued explanations that connected failures to
actionable responses. P12 noted: \textit{``What can be done, how it can be done''} and wanted \textit{``Mitigation can also take into account of the domain/industry of the application''}. Participants asked the judge to look for incidents that describe specific failure mechanisms, trade-offs, and side-effects (e.g., consent processes over time, notification risks) rather than incidents that only provide generic descriptions (P4). This means judges should consider whether incidents describe concrete failure scenarios with enough detail for users to understand what went wrong and how.

\textbf{Expertise}. Judges should evaluate whether incidents match the sophistication level needed for different user profiles. P4 noted that researchers and developers \textit{``have really high standards''} and \textit{``would expect it to give something we don't know,''} while novice users might find basic information helpful. This means judges should favor incidents with complex, non-obvious failure modes when comparing recommendations for expert users (e.g., AI researchers, senior developers), while preferring incidents with clear, foundational risk patterns when evaluating recommendations for novice users (e.g., designers new to AI, junior developers).

\begin{table*}[htb!]
\centering
\caption{Human-Informed Features Ablation Study: Impact of Feature Removal on Model Performance}
\label{tab:ablation_human_informed}
\begin{tabular}{lcccc}
\hline
\textbf{Model Configuration} & \textbf{Validation Loss} & \textbf{MAE} & \textbf{Features Used} & \textbf{Performance} \\
& \textbf{(MSE)} & & & \textbf{Drop (\%)} \\
\hline
Human-Informed Full Model & 0.0190 & 0.0913 & All 21 features + Text & — \\
(All 21 + Text) & & & (8 Original + 13 Human-Informed) & \\
Without Human-Informed Features  & 0.0324 & 0.1016 & Tech, App, Scale, Deploy,  & +70.5 \\
(8 + Text) & & & Output, User, Model, Inter, Text & \\
\hline
\multicolumn{5}{l}{\footnotesize{Performance drop calculated relative to Full Model validation loss}}\\
\end{tabular}
\end{table*}


\begin{table*}[htb!]
\centering
\caption{Learned coefficients and importance changes after expanding the
original model with human-informed features. Bold text indicates features
introduced through the user study.}
\label{tab:coefficients_human_informed}
\begin{tabular}{lcccc}
\hline
\textbf{Feature} &
\textbf{Coefficient} &
\textbf{Original Rank} &
\textbf{Updated Rank} &
\textbf{Rank Change} \\
& $\beta$ (S.E.) & & & \\
\hline
\multicolumn{5}{c}{\textbf{Original Features}} \\
\hline
Deployment Setting
& 0.839*** (0.010) & 4 & 1 & $\uparrow$3 \\

Application Domain
& 0.823*** (0.010) & 2 & 2 & $\rightarrow$ \\

Technology Type
& 0.791*** (0.010) & 1 & 3 & $\downarrow$2 \\

Scale of Operation
& 0.774*** (0.009) & 3 & 4 & $\downarrow$1 \\

Target User Group
& 0.469*** (0.010) & 6 & 7 & $\downarrow$1 \\

Output Type
& 0.405*** (0.010) & 5 & 8 & $\downarrow$3 \\

Interaction Modality
& 0.369*** (0.010) & 8 & 9 & $\downarrow$1 \\

Model Class
& 0.220*** (0.010) & 7 & 13 & $\downarrow$6 \\

\hline
\multicolumn{5}{c}{\textbf{Human-Informed Features}} \\
\hline
\textbf{Harm Categories}
& 0.561*** (0.010) & -- & 5 & \textbf{New} \\

\textbf{Mitigation Cost}
& 0.498*** (0.010) & -- & 6 & \textbf{New} \\

\textbf{Failure Mechanism}
& 0.356*** (0.010) & -- & 10 & \textbf{New} \\

\textbf{Mitigation Feasibility}
& 0.341*** (0.010) & -- & 11 & \textbf{New} \\

\textbf{Regulatory Regimes}
& 0.254*** (0.009) & -- & 12 & \textbf{New} \\

\textbf{Safeguards Present}
& 0.206*** (0.010) & -- & 14 & \textbf{New} \\

\textbf{Vulnerable Groups Affected}
& 0.193*** (0.011) & -- & 15 & \textbf{New} \\

\textbf{Spread Velocity}
& 0.164*** (0.011) & -- & 16 & \textbf{New} \\

\textbf{Data Visibility}
& 0.149*** (0.009) & -- & 17 & \textbf{New} \\

\textbf{Data Sources}
& 0.137*** (0.011) & -- & 18 & \textbf{New} \\

\textbf{Jurisdiction Country}
& 0.087*** (0.010) & -- & 19 & \textbf{New} \\

\textbf{Data Types}
& 0.072*** (0.010) & -- & 20 & \textbf{New} \\

\textbf{Spread Mode}
& 0.006 (0.009) & -- & 21 & \textbf{New} \\

\hline
Text Similarity
& 0.501*** (0.010) & -- & -- & -- \\
\hline
\multicolumn{5}{l}{
\footnotesize
$\uparrow$ and $\downarrow$ indicate changes in relative importance rank
after feature-set expansion.
} \\
\multicolumn{5}{l}{
\footnotesize
*** $p < 0.001$; ** $p < 0.01$; * $p < 0.05$.
} \\
\end{tabular}
\end{table*}

\subsubsection{Integrating and Evaluating Human-Informed Criteria}

We generated the human-informed features from Section \ref{sec:what} for incidents and products, retrained the model, and evaluated their recommendations against the generic baseline using human-informed rationales from Section \ref{sec:how}. The comparison between the original model and the human-informed model is shown in Table \ref{tab:ablation_human_informed}, and the coefficients of the human-informed model are shown in Table \ref{tab:coefficients_human_informed}. Under the human-informed evaluation criteria, the expanded model
was preferred over the text-similarity baseline in 224 of 245
comparisons (91.4\%). The original evaluation yielded an 86.2\%
win rate.


The enhanced evaluation criteria incorporating human study insights produced more nuanced and specific judge rationales. We present examples below.

Enhanced Contextual Analysis and Jurisdictional Consideration: For Shelve (environment variable sharing tool), the judge correctly prioritized GitHub Copilot compliance risks over Sora video generation issues, reasoning: \textit{``Incident A provides insights into compliance risks and intellectual property issues, which are highly relevant given Shelve's handling of sensitive .env files. The jurisdictional alignment with the USA and the focus on developers as the target user group further enhance its contextual fit... Incident B's focus on personal data and multimedia content does not map well to Shelve's operational context.''}

Implementation-Focused Assessment: For Juno (workplace wellbeing), the judge provided detailed operational analysis: \textit{``Incident A highlights systemic data handling issues, including lack of user consent and data sharing with third parties, which are directly relevant to Juno's safeguards and regulatory compliance obligations. The failure mechanism in Incident A (design flaw) is also more actionable for Juno, as it emphasizes the importance of designing systems that prioritize user consent and data transparency.''}

These examples show the revised judge drawing on contextual
criteria such as jurisdiction, affected users, safeguards, and
failure mechanisms when explaining its preferences.


\section{Discussion}
\begin{figure}
    \centering
    \includegraphics[width=\textwidth]{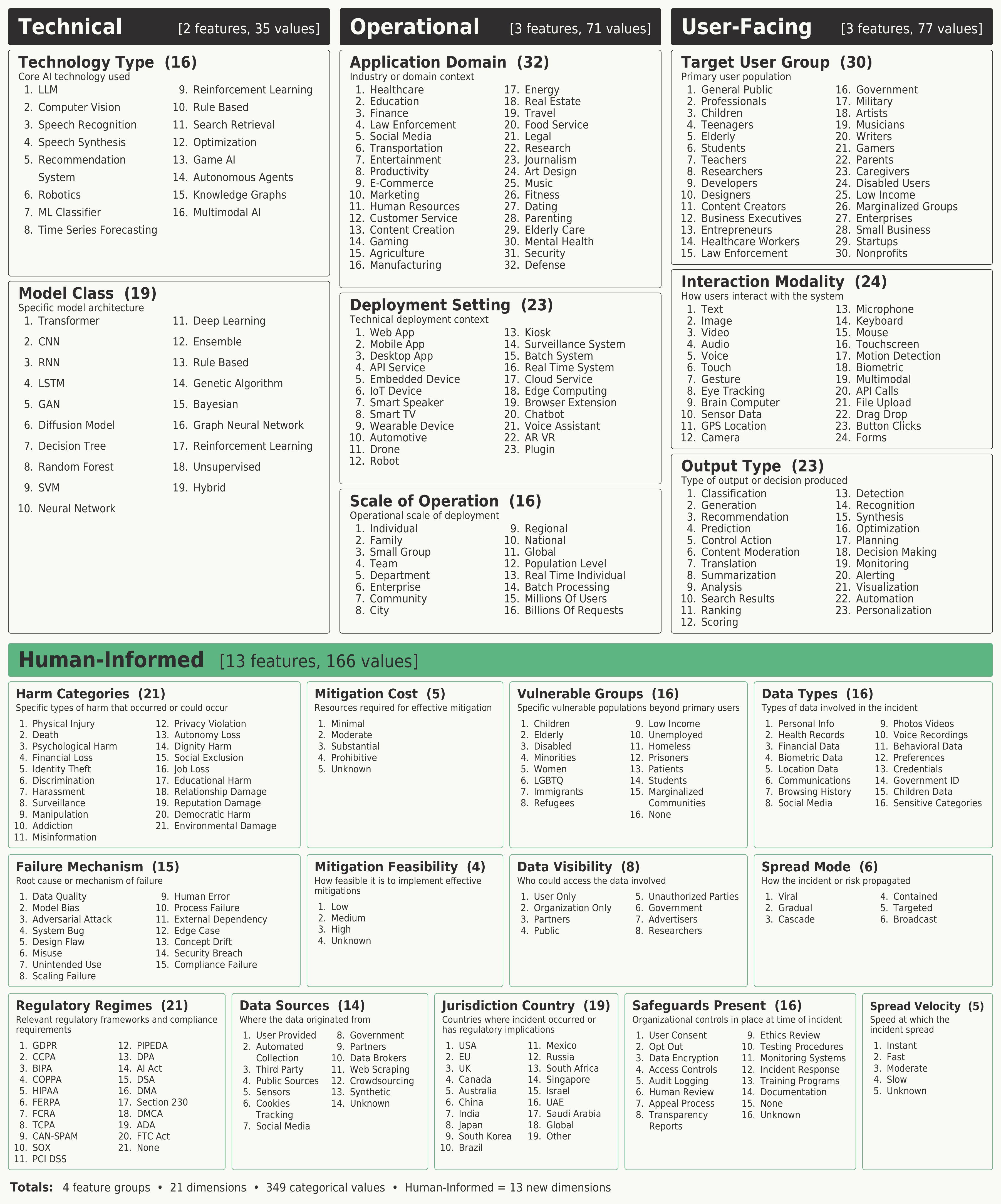}
    \caption{RiskLens Feature Schema of 21 dimensions and 349 values.}
    \Description{A taxonomy chart of the RiskLens feature schema with 4 feature groups, 21 dimensions, and 349 categorical values. Three boxes across the top show the original groups: Technical (technology type and model class, 35 values), Operational (application domain, deployment setting, and scale of operation, 71 values), and User-Facing (target user group, interaction modality, and output type, 77 values). A large green box at the bottom shows the Human-Informed group with 13 new dimensions and 166 values: harm categories, mitigation cost, vulnerable groups, data types, failure mechanism, mitigation feasibility, data visibility, spread mode, regulatory regimes, data sources, jurisdiction country, safeguards present, and spread velocity. Each dimension lists its numbered values.}
    \label{fig:taxonomy}
\end{figure}

Our findings highlight two connected challenges in incident-based AI
risk assessment: identifying historical cases that may inform a new
product, and determining which conditions make those cases applicable.
We discuss what our findings imply for reasoning from historical
incidents, incorporating human judgment into comparison criteria,
and maintaining incident-based risk support.
Figure~\ref{fig:taxonomy} summarizes the expanded feature schema.

\subsection{Human--AI Collaboration for Incident-Based Risk Assessment}

\subsubsection{Incident Transferability as a Contextual Judgment}

Participants' reasoning about safeguards, affected groups, and
deployment conditions (Sections~\ref{sec:what} and~\ref{sec:how})
suggests that the usefulness of a historical incident depends on an
account of why its failure could arise in the target context.
Similarity provides a starting point for this judgment. Its
significance depends on which shared attributes enable a failure
and which differences alter its likelihood or consequences. A case
from another domain may therefore offer a useful lesson, while a
technically similar case may have limited applicability.

This interpretation extends the role of incident evidence in harm
exploration supported by systems such as AHA! and
Farsight~\cite{buccinca2023aha,Farsight}. Historical incidents can
help users formulate conditional questions: which circumstances
enabled this failure, are those circumstances present here, and
what would prevent a similar outcome? The value of a recommendation
consequently includes its capacity to support a reasoned assessment
of applicability. A user who concludes that a recommended incident
is unlikely to transfer because of a relevant safeguard may still
have learned something useful about the product's risks. This
suggests that recommendation quality and the value of the ensuing
assessment are related but distinct concerns for evaluating
incident-based tools.

\subsubsection{Human Feedback Can Revise the Meaning of Relevance}

The study-informed schema expansion illustrates that disagreement
with a recommendation can expose limitations in the criteria used
to produce it. A user may question an incident because an existing
attribute was assigned incorrectly, or because the representation
omits a consideration that matters to their assessment. The latter
requires reconsidering what the system compares. This connects our
findings to responsible-AI research showing how practitioners'
needs and situated organizational conditions shape assessment
practices~\cite{Holstein_2019,Rakova_2021,10.1145/3351095.3372874}.

Human-AI collaboration therefore extends to defining relevance
itself. Translating users' reasoning into reusable criteria makes
contextual concerns available to subsequent assessments, but also
involves choices about which concerns to formalize. A categorical
dimension can preserve a consideration while simplifying the
explanation that originally made it meaningful. For example,
recording safeguard presence does not fully express whether that
safeguard interrupts a particular failure mechanism. The expanded
schema should thus be understood as a revisable representation of
assessment concerns, with human interpretation remaining necessary
to judge their significance in a specific comparison.

\subsubsection{Evidence Growth and Representational Change}

The temporal evaluation and study-informed schema revision address
different aspects of maintaining incident-based risk support.
Newly documented incidents extend the evidence available for
assessment; newly identified criteria change how that evidence can
be interpreted. These forms of change need not occur together. A
repository can grow while continuing to omit a contextual
distinction from its representation, and a useful new criterion
can emerge through reconsideration of incidents already present.

This distinction matters because corpus size alone does not
indicate whether a system can recognize the conditions relevant
to a particular assessment. Maintaining risk support requires
attention both to the coverage of historical evidence and to the
adequacy of the comparison criteria. RiskLens offers an initial
example of treating these as separately revisable components.
Further research can examine when additional cases address an
assessment gap and when the gap instead calls for changes to the
representation or evaluation criteria.

\subsection{Design Implications}

\subsubsection{Supporting Connected Evidence and Reasoning}

Participants' reported preference for combinations of incidents,
scenarios, and mitigations (Section~\ref{sec:combined feature})
suggests that tools should support movement between evidence,
possible consequences, and planned responses. This is consistent
with sensemaking research that connects gathering evidence with
developing and revising interpretive
structures~\cite{russell1993cost,pirolli2005sensemaking}.

Future tools should help users move between these components and
inspect their connections. For example, users could trace a
generated scenario to the incident evidence and product attributes
that motivated it, then examine the assumptions behind a proposed
mitigation. Interfaces should distinguish documented events,
generated possibilities, and editable risk assessments so that
users can assess the basis of each claim. Supporting these
connections could help users develop an assessment while retaining
access to the evidence needed to question it.

\subsubsection{Making Context and Limits of Applicability Inspectable}

Participants' attention to contextual conditions
(Sections~\ref{sec:what} and~\ref{sec:how}) motivates interfaces that
highlight consequential differences alongside feature matches and
invite users to explain whether a difference limits transfer. They
could also distinguish information explicitly provided in a
product description or incident report from inferred or unspecified
attributes. This would allow users to identify where a
recommendation depends on an assumption and supply contextual
knowledge that the available descriptions do not contain. The aim
is to support examination of the conditions under which a risk may
arise and the practical options for addressing it.

\subsubsection{Visualizing Possible Risk Propagation}

P5's request for clearer representations of how failures spread to
downstream harms suggests a direction for supporting temporal and
relational reasoning (Section~\ref{sec:what}). Timelines or pathway
diagrams could connect an initiating failure with affected
stakeholders, subsequent consequences, and possible intervention
points. Such representations could help users examine how an
incident might unfold in their product context and where a
mitigation might interrupt that progression.

These visualizations should distinguish documented relationships
from hypothesized extensions to the target product. Users should
be able to inspect and revise the assumptions connecting stages
of a proposed pathway. This would support examination of a risk
scenario; explaining the ranking score would additionally require
showing how the model's comparison signals contributed to the
recommendation. RiskLens did not implement or evaluate propagation
visualizations, and their effects on understanding and assessment
quality remain questions for future work.

\subsubsection{Accommodating Different Forms of User Feedback}

Participants varied in the specificity of guidance they sought.
For example, P4 emphasized the value of insights beyond familiar
risk considerations. Future feedback interfaces could accommodate
such differences by offering several ways to contribute:
correcting a feature assignment, explaining a contextual mismatch,
comparing alternative incidents, or expressing uncertainty about
applicability.

These options would allow users to contribute without requiring
everyone to provide a detailed technical diagnosis. Follow-up
questions could ask what information would resolve uncertainty or
which aspect of an incident made it useful. Future studies should
examine how these forms of feedback vary with users' experience
and assessment goals, and which contributions support reliable
revisions to the comparison criteria.

 \subsection{Limitations}
Our study involved 21 AI-familiar participants in a single session,
limiting conclusions about other user groups, sustained adoption,
and long-term changes in assessment practices. Without a control
condition or independent ratings of plan quality, the observed
changes reflect participants' self-assessments and articulated
reasoning; they cannot be attributed to the recommender alone.

RiskLens relies on ProductHunt listings and publicly documented
AIAAIC incidents, whose coverage and descriptive detail constrain
its recommendations. LLM-extracted features may also be inaccurate
or incomplete. The human-agreement
check ($\kappa = 0.71$, $n = 50$) provides partial validation, but
broader assessment remains necessary.

Our evaluations cover one temporal holdout and one study-informed
schema revision, rather than sustained adaptation across repeated
updates. The findings therefore do not establish continuous
validity or automatic improvement as the corpus grows. Comparisons
with a text-similarity baseline also do not establish the relative
effectiveness of complete workflows such as Farsight or AHA!.

\bibliographystyle{ACM-Reference-Format}
\bibliography{references}

\end{document}